\documentclass[journal]{IEEEtran}
\usepackage{xcolor,soul,framed} %,caption
\usepackage{graphicx}
\usepackage{amsmath}
\usepackage{array}
\usepackage{mdwmath}

\usepackage{mdwtab}
\usepackage{eqparbox}
\usepackage{url}
\usepackage{lineno,hyperref}
\usepackage{amsmath,amssymb}
\usepackage{booktabs}
\usepackage{makecell}
\usepackage{multirow}
\usepackage{multirow}
\usepackage{tabu,bm}
\usepackage{color}
\usepackage{lipsum,mwe,cuted}
\usepackage{stfloats}
\usepackage{cite}
\usepackage{enumitem} 
\usepackage{diagbox}
\usepackage{subcaption}
\title{Computing Networks Enabled Semantic Communications}

\author{Zhijin Qin, Jingkai Ying, Dingxi Yang, Hengjiang Wang, and Xiaoming Tao
\thanks{Zhijin Qin, Jingkai Ying, Dingxi Yang, and Xiaoming Tao are with the Department of Electronic Engineering, Tsinghua University, Beijing, China. (e-mail: qinzhijin@tsinghua.edu.cn;taoxm@tsinghua.edu.cn) }
\thanks{Hengjiang Wang is with China Mobile Group Device Co., Ltd. and Tsinghua University, Beijing, China. (email: wanghengjiang@cmdc.chinamobile.com) }
}

\begin{document}

\maketitle

\begin{abstract}
Semantic communication has shown great potential in boosting the effectiveness and reliability of communications. However, its systems to date are mostly enabled by deep learning, which requires demanding computing resources. This article proposes a framework for the computing networks enabled semantic communication system, aiming to offer sufficient computing resources for semantic processing and transmission. Key techniques including semantic sampling and reconstruction, semantic-channel coding, semantic-aware resource allocation and optimization are introduced based on the cloud-edge-end computing coordination. Two use cases are demonstrated to show advantages of the proposed framework. The article concludes with several future research directions.
\end{abstract}

\section{Introduction}
The sixth generation (6G) networks are being investigated to address diverse user requirements effectively and ensure a high quality of experience (QoE) for everyone, adapting to the needs of various user segments. 
An artificial intelligence (AI)-driven architecture will support the vision of an open, flexible, and decentralized network, and will play a key role in providing everyone-centric customized services.
Semantic communication, as the second level of communication, takes semantic information ignored by the bit pipeline into consideration, showing great potential for a more efficient and intelligent communication paradigm. 
This enables a more efficient use of the network by transmitting the most relevant and pre-processed information  with advanced AI technologies.
Semantic communication powered by AI in 6G is to be both foundational and transformational, enabling the network to be more dynamic, responsive, and capable of handling the intricate requirements of 
emerging applications with huge data volume, such
as Extended Reality (XR) and holographic video.

Pioneering works in semantic communications followed the path of typical information theory, trying to develop semantic theories based on logical probability~\cite{bao2011towards}. However, it is still in infancy and not yet able to provide theoretical limits and analytical guidance for semantic communication systems as Shannon information theory did. Thanks to the profound advancement of deep learning (DL), the powerful nonlinear fitting and feature extraction capabilities of deep neural networks (DNN) make the realization of semantic communication systems possible. DL-enabled semantic communications with significant performance gain and potential have drawn tremendous attention for the design  of system architectures, loss functions, performance metrics, training algorithms and so on~\cite{qin2021semantic}. A variety of well-known and high-performing DNN models, such as convolutional neural network (CNN), generative adversarial network (GAN) and Transformer, have been applied to semantic communication systems. Despite their excellence in system performance, DNN models induce an enormous computational burden on communication devices. Computing capability becomes the bottleneck of DNN enabled communication systems, especially semantic communication systems.

However, end devices often lack the necessary computing power to support semantic communications. The efficient integration of communication networks and computing resources is an important issue to be addressed to advance practical applications of semantic communications. Cloud computing offers abundant computing resources through the network, but it suffers from drawbacks such as high transmission latency and significant spectrum resource usage. Edge computing leverages the distribution of computing resources towards network edges, such as base stations or gateways closer to end-users, enabling faster interaction with users. But individual edge computing nodes have limited resources and lack effective coordination among end devices. Edge computing devices and cloud computing servers led to issues such as low resource utilization and imbalanced load. Thus, the concept of computing networks is proposed to address these challenges. \emph{Computing network} is a novel information infrastructure, revolving around computing power, which can connect various computing resources through networks and provide integrated services by coordinating business demands based on computing network states. It connects previously isolated cloud, edge, and end computing nodes via the network, facilitating efficient resource coordination and scheduling. The computing network provides computing power support for semantic communications and fulfills its requirements in terms of response speed and spectrum resource utilization through end-edge-cloud collaboration.

In this article, we propose a framework of  computing networks enabled semantic communication, which differs from existing ones by introducing the connected computing. The essential features of computing networks enabled semantic communication systems come from twofold: First,  semantic communication system could fully exploit the available computing resources to support semantic processing modules, such as semantic sampling and joint semantic coding, with the help of computing networks. Second, the semantic communication components could function well in a manner adaptive to available computing resources.

The rest of this article is organized as follows. Section II provides a brief view of computing networks and semantic communications. In Section III, the overall architecture and key techniques of the computing networks enabled semantic communication are detailed. Two use cases are demonstrated in Section IV. Section V concludes this article with outlooks.

\section{Overview of Computing Networks and Semantic Communications}
In this section, we first introduce the basic concept, key technologies and critical challenges of computing networks, which could facilitate the implementation of semantic communications. Subsequently, several potential frameworks of DL-enabled semantic communications are presented.

\subsection{Computing Networks}
\begin{figure}[t]
\centering
\includegraphics[width = 3.2in]{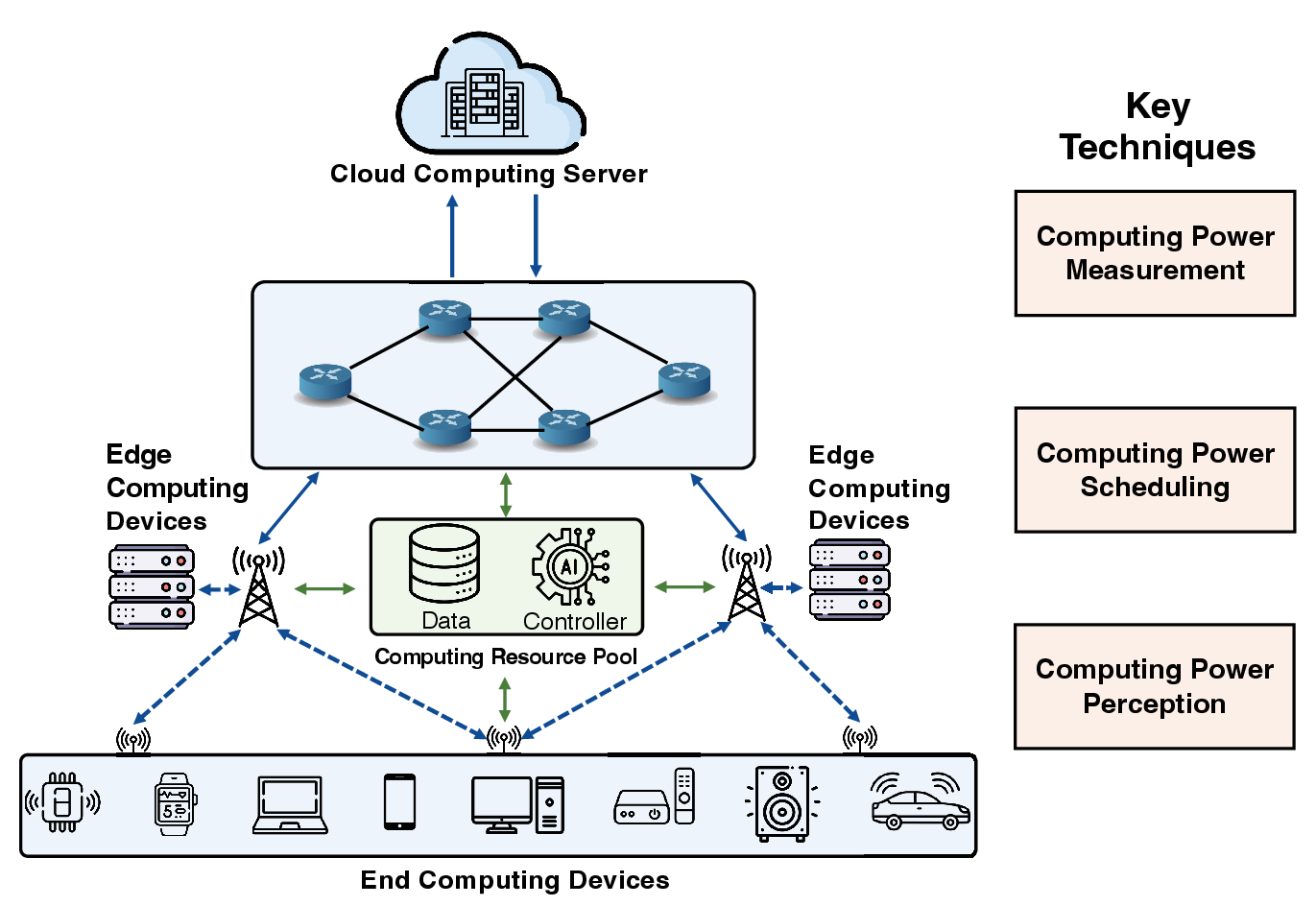}
\caption{A computing network with its key techniques.}
\label{computing_network}
\end{figure}

The concept of computing network is to interconnect widely distributed cloud, edge, and end computing nodes. It abstracts and pools heterogeneous computing, storage, and network resources into a computing network resource pool. The computing network continuously monitors the availability status of resources in the network and dynamically allocates computing resources, including computing power, based on user demand. It also partitions and offloads tasks accordingly. By doing so, it satisfies the computing demands of applications, improves the utilization of computing resources, and enhances the overall user experience. Specifically, the integration of hierarchical computing paradigms with artificial intelligence (AI) can promote the rapid development of AI applications~\cite{duan2023}. A generalized architecture of computing networks with its key techniques is illustrated in Fig.~\ref{computing_network}.

The key techniques of the computing network include computing power measurement, computing power perception, and computing power scheduling. Computing power measurement, as the foundation for computing power perception and scheduling, requires the establishment of a universal quantification model to assess the computing power demands of users and the status of the computing power resource pool. In \cite{zeng2021Energy}, the computing speeds of the CPU and GPU are calculated based on their clock frequencies respectively. Workload division and allocation can be conducted based on the computing power measurement of heterogeneous computing infrastructure.

Computing power perception refers to the real-time comprehensive awareness of the computing resource status and task demands, ensuring computing power scheduling. In \cite{lin2019comm}, the edges actively monitor the available communication and computation resources, enabling them to estimate the end-to-end latency and achievable data rate accordingly. Furthermore, the system load is estimated before accepting connections to the edges, ensuring that the overall system load remains within the limits of the edge capabilities.

Computing power scheduling involves partitioning and offloading tasks to appropriate computing nodes based on task requirements, aiming to enhance user experience and improve the utilization of computing resources. In \cite{deep_reinforce}, a policy gradient deep reinforcement learning approach was employed for decentralized computation offloading. For tasks with large computation volumes, it is necessary to divide the computational tasks and offload them to multiple computing nodes. \cite{post} proved the existence of generalized Nash equilibrium for the parallel offloading of divisible tasks problem and developed a corresponding distributed task offloading method.

However, the research on computing networks still faces numerous challenges. First, due to the issue of heterogeneous software and hardware in edge and end devices, unified resource management poses difficulties. Second, the real-time and rapid changes in the computing status of network nodes present certain challenges to real-time computing power perception. Lastly, issues related to data sharing and data security between computing nodes require further exploration.

\subsection{Deep Learning enabled Semantic Communications}
Joint semantic-channel coding (JSCC) approaches are widely utilized in the design of semantic communications. Shannon’s source-channel separation theorem enables us to design source coding and channel coding independently without losing optimality, however, infinite block-length is required. Some recent results implied that joint source-channel coding can outperform separate schemes, especially in the short block-length regime~\cite{JSCC_text}. DNN-aided joint source-channel coding is a method to realize joint design. The idea of implementing coding models by DNN can be naturally incorporated into DL-enabled semantic communication systems which emphasize extracting and compressing semantic information. By considering semantic information, channel conditions and tailored performance metrics collaboratively, DL-enabled semantic communications with JSCC can achieve exceptional end-to-end performance.

In\cite{DeepSC}, the proposed DeepSC consists of a semantic encoder and a channel encoder in the transmitter, and corresponding decoders in the receiver. All modules were implemented by neural networks. After training the mutual information estimation model, the whole system was trained jointly to preserve semantic information while compressing data. In\cite{DeepSC_Lite}, DeepSC was deployed in an Internet-of-Things (IoT) network where the cloud/edge platforms equipped with high computing capabilities are used to train and update DL models while  distributed IoT devices are used to extract and upload semantic features. Based on \cite{DeepSC_Lite}, computing task offloading problems in semantic communication systems was further studied, and a multi-agent reinforcement learning algorithm was proposed to coordinate the resources of communications and computing\cite{DeepSC_Resource}. DeepSC performed the JSCC schemes in a designing separately while training jointly manner, which owns convincing interpretability. In \cite{Wireless_Video}, focusing on video conference, a semantic encoder was utilized to extract facial keypoints and a Hybrid Automatic Repeat Request (HARQ) scheme equipped with a semantic error detector was introduced to cope with impairments brought by the wireless channel. It is a novel way to implement JSCC designs. Moreover, a DL-enabled semantic communication system with a data adaptation module added before the JSCC encoder, which can keep the system functioning properly when required semantic information changes, was proposed in \cite{Task_Unaware}.      

While the above works show the popularity of JSCC in current designs of semantic communication, there are some other approaches to realize semantic exchange. In \cite{LLM}, the powerful Large language model (LLM) has been utilized to quantify the semantic importance. Specifically, whether a word in a sentence is important or not is determined by the LLM model. Equipped with a cross-layer manager, the LLM model can incorporate into existing communication systems and perform semantic-aware power allocation. Some researchers pursue System 2 semantic communications, which owns high-level cognition and connections between knowledge, agency and reasoning. In \cite{System2}, inspired by the real-life phenomenon that humans can communicate more efficiently if the listener can reason from the speaker's context, the semantic-native communication system employed a virtual listener at the transmitter side and a virtual speaker at the receiver side for contextual reasoning. Although numerous examples have demonstrated the superiority of DL-enabled semantic communications, sufficient computing resources are required to bring out their excellent performance. This has prompted us to conduct research on the combination of computing networks and semantic communications. 

The complexity and the requirements for real-time processing of semantic tasks make it heavily reliant on substantial computing power to perform effectively. Unlike traditional communication resources, computing power can be more ubiquitous. Devices at the end and edge are capable of providing computing power with low latency. The computing network is designed to connect and manage this computing power, which could be seen as one of the objectives rather than a constraint. The fusion of computing networks and semantic communication goes beyond merely allocating computing power for semantic tasks; it aims to build networks for comprehensive planning of the computing power to efficiently facilitate semantic communication.

\section{Computing Networks Enabled Semantic Communications}
As aforementioned, most existing DL-enabled semantic communication designs require heavy computing resources. Therefore, it is intuitive and imperative to deliberate on computing when designing semantic communication systems. In this section, we first illustrate the framework  of the computing network enabled semantic communication system with its elemental modules. Subsequently, the enabling technologies of the proposed framework are detailed.

\subsection{Proposed Architecture of a Cloud-Edge-End Computing  enabled Semantic Communication}
\begin{figure*}[t]
\centering
\includegraphics[width =0.85\linewidth]{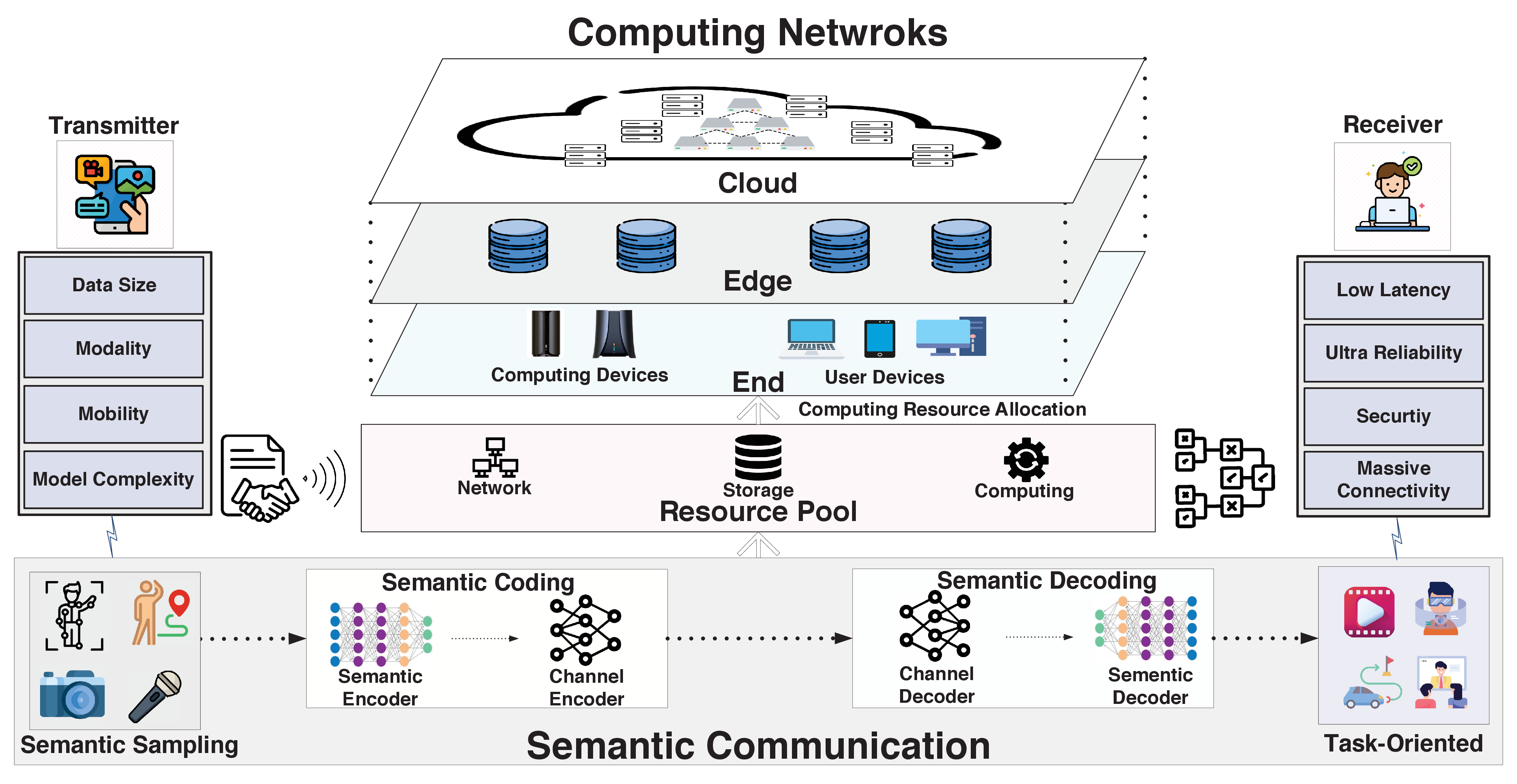}
\caption{The  proposed architecture of cloud-edge-end computing enabled semantic communication system.}
\label{SystemModel}
\end{figure*}

5G networks focus on the average user and tend to deliver satisfactory performance for standard use cases.
% such as enhanced Mobile BroadBand (eMBB), Ultra Reliable Low Latency Communications (URLLC), and massive Machine Type Communications (mMTC). 
Nonetheless, the ambition for 6G technology is to shift towards a highly personalized service approach. This new architecture aims to cater specifically to the varying, complex, and evolving service needs associated with the diverse tasks.
To achieve task-level service, efficiently utilizing AI is essential. 
Two aspects of 6G with native AI support are mostly concerned, which are AI4Net, using AI technology to improve network performance, and Net4AI, providing enhanced AI services by the network. 
Motived by the above, we proposed a cloud-edge-end computing network enabled semantic communication architecture.
Semantic communication serves as the core to providing everyone-centric customized services with intelligent, meaning-based connections.
The cloud-edge-end computing network serves as a platform to provide efficient computing power, increased coverage, and enhanced AI capabilities, enabling distributed and collaborative semantic services.
The architecture of the proposed cloud-edge-end computing-enabled semantic communications is illustrated in Fig. \ref{SystemModel}. The semantic communication powered AI4Net modules of the proposed architecture are as follows:

\textbf{Semantic Sampling:}  \textit{Semantic sampling} is an emerging technique employed at the transmitter based on programmable sensors to lower the network traffic assisted by the computing resource. Instead of taking samples based on traditional rules such as random sampling or uniform sampling in frequency or time domains, semantic sampling takes samples by considering their semantic meaning or relations to the task execution at the receiver. Particularly, semantic sampling will use more samples for semantic contents harder to be sensed, such as moving targets (human, animals, and cars), non-sparse contents (colorful items with rich details), or structurally-complicated objects (artistically-designed buildings) while reducing samples for other semantic contents. Semantic sampling will also intelligently prioritize and process data that is most semantically relevant or valuable for various tasks. By doing so, less amount of data is generated, processed, and transmitted, so that to provide a more intelligent, context-aware approach to data processing, especially in environments where spectrum resources are limited. This technique can be particularly advantageous in scenarios where multiple devices generate a vast amount of data, but not all of it has dynamic/non-sparse/architecture-complicated objects and is equally important or useful for tasks at the receiver. 

\textbf{Computing at the Transmitter:} For supporting semantic sampling, several factors that substantially influence computing requirements are critical for efficient network design and operation. First, the size of the data being processed at the transmitter plays a significant role in computing resources. Larger datasets demand more computing resources for tasks like data storage, transfer, and processing. Second, the modality of the data, whether it is structured or unstructured, text-based or media-rich, also determines the computational complexity. Generally, multimodal, unstructured, or media-intensive data often requires more powerful computing capability for efficient processing. Third, for networks with high mobility, the dynamic nature of the network topology increases the computation requirements for tasks, such as routing, resource allocation, and handover management. Finally, the model size of neural networks or the complexity of algorithms is also a significant factor. Usually, complex models require substantial computing power for training and inference. By considering these factors, computing networks could better predict and manage the computing power required in different scenarios.

\textbf{Semantic Communication:} Following the semantic sampling operation, semantic communication allows devices to transmit data based on its semantic meaning, which could improve the transmission efficiency significantly. This approach requires the complex encoder and decoder that transform semantic samples into semantic symbols and then reverts them back into a structured format, respectively. This codec, mainly realized by JSCC approaches enabled by DL, may come with high computational requirements, making them challenging to deploy on resource-constrained devices. However, through the use of advanced cloud-edge-end computing strategies, the computational tasks of semantic communication can be efficiently distributed and managed.

\textbf{Task Execution at the Receiver:} Semantic communication, which revolves around the delivery of data based on meaning, is highly dependent on the specifics of the task at hand. Different tasks require different types of semantics, with varied significance levels and formats. Therefore, having a task-oriented design ensures that the most appropriate and meaningful semantics are transmitted efficiently for each specific task. For the cloud-edge-end coordination computing network, task-oriented design aids in optimizing network performance and resource allocation. As different tasks come with diverse requirements in terms of latency, reliability, security, and connectivity, a design that takes them into account could better prioritize and allocate network resources, enhancing overall efficiency and performance.

The computing network powered Net4AI modules of the proposed architecture are as follows:

\textbf{Cloud-edge-end Architecture:} Cloud-edge-end architecture is the core of computing networks. The cloud is usually with powerful, centralized servers with rich computing resources and storage capabilities, which is excellent for processing large-scale and complex tasks. However, due to its distance from the end-user, cloud computing may introduce extra transmission latency, making it inapplicable for time-sensitive applications. Edge computing is positioned closer to end-users or data sources, and these edge devices provide a type of middle-level computing capability, capable of executing tasks with lower latency than the cloud. They serve as a bridge between the cloud and the end-users, managing real-time or near-real-time data processing, reducing network congestion and latency, and providing localized decision-making capabilities. The end refers to the user devices or end computing devices in the network. These devices are often with lower computing capabilities but crucial for tasks requiring immediate and local processing. This tri-layered approach allows for dynamic and intelligent allocation of computational tasks, ensuring that each task is handled at the most appropriate layer to satisfy its requirements, resulting in an optimized and efficient computing network. 

\textbf{Resource Pool:} A resource pool in the context of computing networks is a comprehensive collection of resources that are essential for network operation and performance, which usually includes elements such as network condition, storage capability, and computing power. The network condition, including factors such as transmit power, bandwidth availability, and latency, plays a crucial role in data transmission and affects how computing resources are allocated. Storage refers to the capability to store data, whether temporarily or long-term, across various devices and locations within the network. The computing power represents the processing capabilities of devices in the network, which is vital for executing tasks and computing resource allocation.

\subsection{Cloud-edge-end Computing  enabled Semantic Sampling, Coding and Reconstruction} 
\begin{figure*}[t]
\centering
\includegraphics[width=1.8\columnwidth]{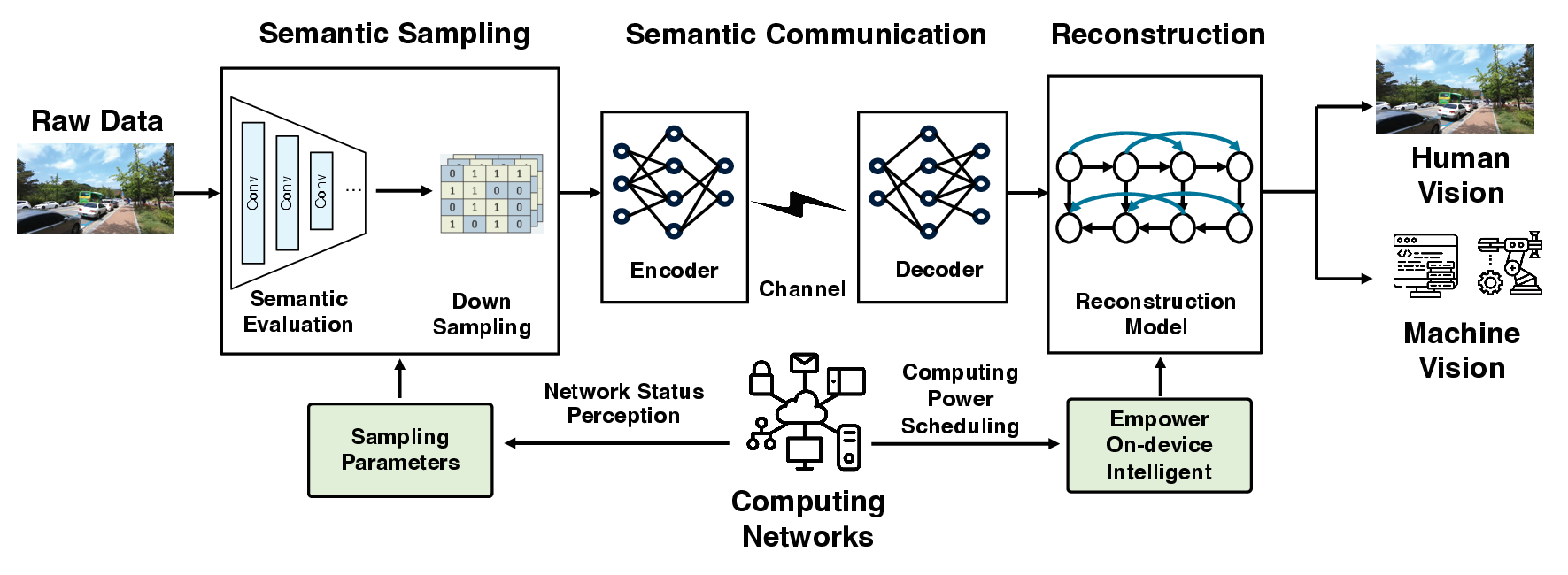}
\caption{The structure of computing networks enabled semantic sampling, coding and task-oriented reconstruction.}
\label{sampling}
\end{figure*}

This part introduces the enabling techniques for cloud-edge-end coordinated computing in semantic sampling, coding and task-oriented reconstruction as shown in Fig.~\ref{sampling}.

\subsubsection{Computing Networks enabled Semantic Sampling and Task-oriented Reconstruction} 
The massive amount of data collected by sensors poses a significant burden on both the network and computing resources. However, these raw data often contain a large amount of redundancy. To address this issue, semantic sampling techniques can be employed to selectively sample data based on their semantic information, thereby reducing the costs of communication and computation. 

Semantic sampling, which consists of two stages, is performed by intelligent sensors. In the first stage, the intelligent sensors are programmed to selectively activate specific sensing elements to reduce the amount of sensed data. The selective sampling process is guided by the specific task requirements and the the feedback of task performance from the receiver. In the second stage, intelligent sensors preprocess and downsample the raw data to further decrease the data volume. Limited computing power prevents common sensors from performing semantic sampling.

Take the video transmission, which occupies the majority of traffic in the network, as an example. For specific tasks, certain sensor elements will be deactivated while task-related data will be sampled. The strategy is based on historical data and the feedback of task performance from the receiver. To further reduce data redundancy within video frames, spatial sampling can be performed. Specifically, a semantic-based low-pass filter can be designed to minimize the loss of information in the downsampled low-resolution frames and anti-aliasing at the same time. The parameters of the filter are jointly trained with the reconstruction module. The core idea of semantic sampling is to discard data that has low semantic relevance while ensuring the remaining data contains complete semantic information. Therefore, in addition to videos, semantic sampling is also applicable to multimodal data such as images, audio, text, and so on.

However, it is hard for sensor devices with limited computing power to train semantic sampling models. So it is necessary to transfer the data collected by the sensor devices to other distributed computing nodes so as to complete the training and updating of the models. These models are then deployed on intelligent sensors to enable semantic sampling through the computing network. 

The parameters of semantic sampling need to be determined based on the real-time state of the computing network. Particularly, the computing network needs to perceive the network state, the requirements of the tasks, and available computational resources at the receivers, so as to guide semantic sampling. For example, if the computing network detects network congestion and the task at the receiver has low requirements for the quality of received data, only most important semantics are taken and the fine will be discarded by semantic sampling. In addition, the computing network also needs to consider whether the receiver has sufficient computational resources. Otherwise, after sampling, other nodes near the receiver will be scheduled to pre-allocate computational resources for the receiver to execute DL-enabled tasks, such as generative reconstruction, classification, and object identification.

Continuing the video transmission example shown in Fig.~\ref{sampling},  with decoded semantic samples at the receiver, we will reconstruct the semantically sampled videos to restore their original resolution. For instance, video super-resolution techniques can utilize features from received frames to reconstruct using bidirectional recurrent neural networks. And optical flow estimation and deformable convolution techniques are employed jointly for refinement. The aforementioned reconstruction processing requires significant computing resources. Although the computing capability of receivers has been considered in the sampling phase, the transmitter cannot offer a precise assessment of the computing network state at the receiver side. So a quick measurement of the computing network state is needed to determine whether a computing power scheduling should be performed. When end devices face challenges in executing intelligent enhancement tasks independently, it becomes necessary to offload these tasks to the computing networks for collaborative processing. 

\subsubsection{Computing Networks enabled Joint Semantic-Channel Coding} 
JSCC consists of semantic encoding and channel encoding at the transmitter and corresponding decoding modules at the receiver, which is the core of  DL-enabled semantic communications. The transformer structure or multiple convolutional layers are used in the semantic encoder to extract semantic features of text, speech, image and video, requiring a great amount of computing resources. Therefore, JSCC is difficult to deploy directly on distributed end devices with limited computing capability. The practical applications of semantic communications are hindered. 

To solve the problems, idle distributed computing resources and powerful servers in cloud/edge platforms should be exploited. The computing networks enabled joint semantic-channel coding exploits computing resources in the resource pool to support model inference and training. Such a JSCC approach also contains hierarchical models to match the cloud-edge-end structure of computing networks. The comprehensive models are deployed on the cloud or the edge, while they are compressed by pruning and parameter quantization before deploying on end devices with acceptable performance degradation. Complex tasks such as semantic encoding and semantic decoding will be decomposed and performed at different computing devices. To support such a decomposition,  task offloading and computing resource scheduling should be performed based on the status of computing networks, which will be detailed later of this subsection.

With the computing network, semantic-channel coding schemes are designed to be computing-adaptive. Compared with most existing semantic-channel coding, the main difference is that we fold the status of computing networks, as an input of encoder and decoder, into the coding schemes. From a long-term statistical perspective, the status of computing networks tends to exhibit certain spatial-temporal distributions. For instance, in office buildings during working days, there are relatively few idle computing resources and network congestion might be a common issue. In such a case, the coding schemes could be more conservative with less number of extracted semantic features  and simpler channel coding designs. We  train the JSCC model with preset status and coding results to inject computing elements into the JSCC design, developing computing-adaptive joint coding schemes.

\subsection{Joint Computing-Transmission Resource Allocation in Semantic-aware Networks}
This part introduces the task-oriented offloading, and new performance metrics for semantic-aware networks.
\subsubsection{Task-oriented Offloading} 
As a typical scenario of the computing-enabled transmission system, semantic communications provides an advanced task offloading scheme itself. Unlike conventional cloud/edge computing systems where the end devices offload the source data to cloud/edge server, semantic-aware task offloading systems extracted the task-oriented semantics and  offload them to the computing server, i.e., cloud or edge. The issue of the task offloading systems lies in the poor computing capabilities of the end devices. The limitation of the end device inhibits the further development and application of the DL model on the local side. By leveraging the computing networks, the high computing complexity of local devices can be reduced or offloaded, converting the local computing requirement to thE cloud or edge servers with idle computing resources, with low transmission overhead. 

\begin{figure}[t]
\centering
\includegraphics[width=\columnwidth]{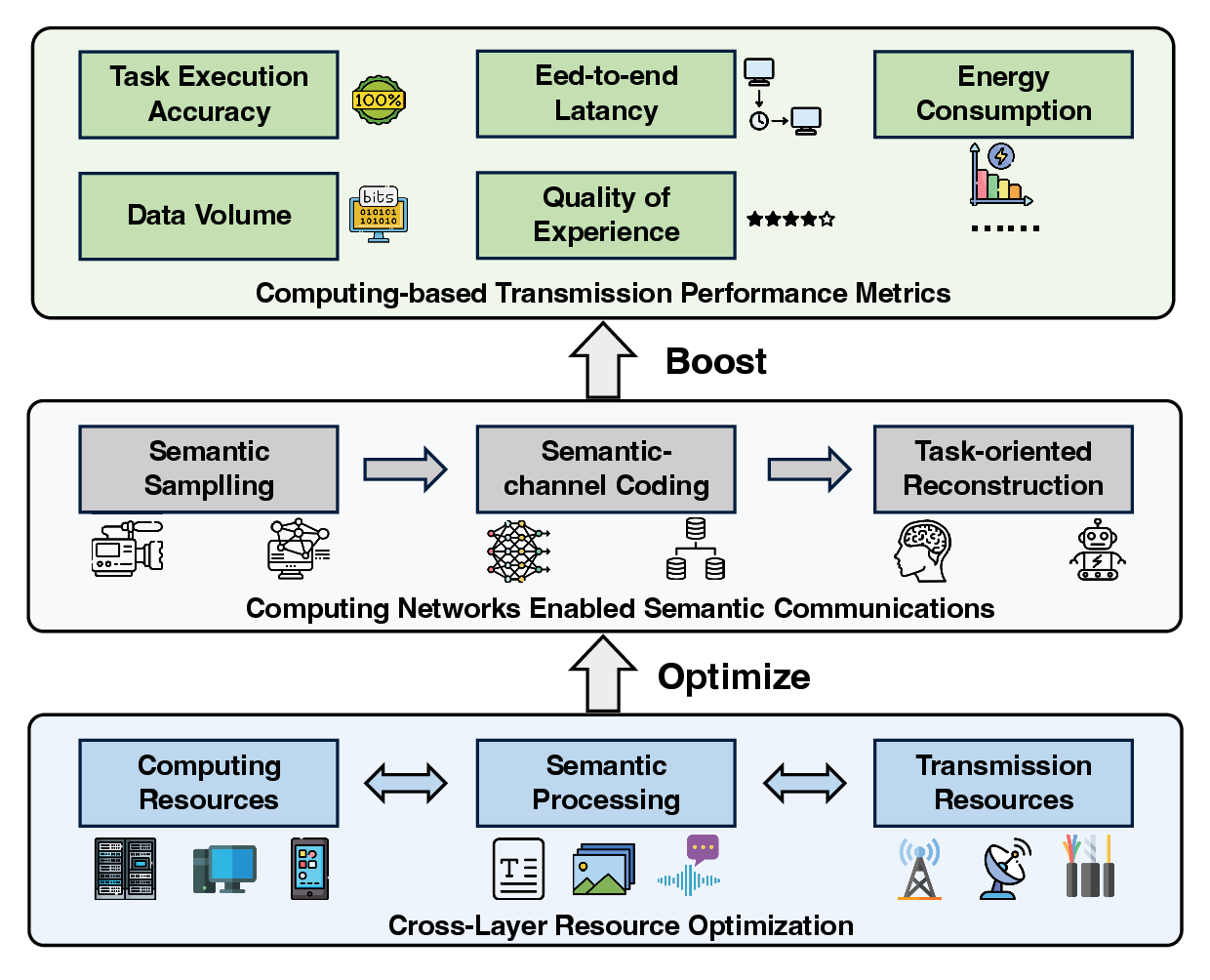}
\caption{Cross-layer resource optimization of the computing-enabled transmission.}
\label{optimization}
\end{figure}

However, to coordinate the resource of the cloud-edge-end, a cross-layer resource optimization framework need to be designed. As shown in Fig.~\ref{optimization}, the key resources include the transmission-related resources, e.g., transmit power, channel allocation, user pairing, etc, and the computing-related resources, including computing power, computing frequency, computing offloading policy, user association, etc. Nevertheless, these resources are heterogeneous and may not be accessible by the public. To coordinate the resource for computing-based transmissions, especially for semantic communications, enabling techniques should be designed, which will be detained by the use case in Section IV.

\subsubsection{Performance Metrics and Strategies for Resource Optimization}  
Due to the heterogeneity of the end-side computing network and end devices, a unified indicator that reflects the QoE should be designed. Such an indicator should consider factors such as transmission overhead, computing cost, total execution delay, total energy consumption, task execution performance, to name a few. 

Compared to the conventional edge computing system the source data of the task needs to be transmitted to the servers by bit stream, the semantic communications extract the task related information to semantic features at the end users, and the users only offload the extracted semantic features to the servers. Therefore, we need to consider the semantic rate instead of the bit rate of the offloading process. To measure the semantic task execution latency and the energy consumption, the costs of the end users, the offloading process, and the servers should be jointly considered. Moreover, for different devices and computing networks, the QoE preference could be various. Moreover, the dynamic QoE poses a higher requirement on the robustness of the resource optimization algorithm.

Basically, computing resources could be allocated by the central cloud, or accessed by end devices in a distributed manner. The centralized resource optimization could coordinate and aggregate the global resource, so as to strengthen the overall plan of the global resource for the whole computing network. On the other hand, the computing resources of end devices are not always available to the public, and centralized optimization could increase the computation pressure of the cloud significantly. A promising solution is leveraging the distributed resource allocation approaches. For the fully competitive scenarios, the game theory algorithm could be utilized to achieve the Nash Equilibrium point, otherwise, the multi-agent machine learning algorithms could be leveraged to motivate the participation and collaboration among the cloud-edge-end devices. 

\section{Use Cases}
In this section, we will demonstrate the advantages brought by the computing networks enabled semantic communications. In particular, two use cases are provided.

\subsection{Computing Networks enabled Video Super Resolution}
We propose an end-cloud computing enabled video transmission system to solve the limited spectrum resource problem in video transmission using video super-resolution. Particularly, we focus on the scenario involving end-side cameras as the sender, transmitting surveillance videos to the cloud. 

First, the video to be transmitted is spatially downsampled at the end side. Specifically, after deconstructing the source video into a frame sequence, a 4× bicubic downsampling  is performed in the spatial dimension for each frame to acquire the corresponding low-resolution (LR) frame sequence. Second, key frames are extracted from the source video. Apart from selecting key frames in a sparser manner, a key frame selection strategy based on the content of the video is designed to maximize the utilization of key frame information. Subsequently, a redundant frame elimination module is adopted to detect and remove redundant frames from the LR frame sequence, further reducing the data volume and improving the efficiency of the system. The detection of redundant frames is based on two criteria, which include the mean square error (MSE) between successive frames and the MSE of the detected motion regions. Then conventional source coding is performed on both the LR video and the key frames and the overall code stream is transmitted to the cloud.

The cloud server decodes the received code stream to obtain the LR video and key frames. Then they are input into the proposed video super-resolution model to generate the reconstructed HR frames.
First, feature extraction is performed on the input LR frame sequence and HR key frames. Then, the extracted features are input into the recurrent neural network (RNN) propagation module, which consists of four layers. Within each layer, the first-order propagation is performed in chronological order, and at the same time, the key frame features are directly propagated to each LR frame. Then the refined features are propagated downwards layer by layer. Finally, upsampling is performed using multiple convolutional layers and a pixel shuffling layer to obtain the reconstructed frame sequence. 

To obtain the complete HR reconstructed video at the cloud platform, according to the received redundant frame position information, the HR frames corresponding to the redundant LR frames which are removed before transmission are reconstructed by copying the non-redundant frame. 
Different key frame intervals result in varying bits per pixel (bpp) and experiments are conducted on the cases of different fixed key frame intervals. Fig. 5 illustrates the relationship between the reconstruction  quality and the data volume under different key frame intervals. As the key frame interval increases, the performance degradation rate of KA-VSR is considerably slower than that of NeuriCam, which demonstrates the performance advantage of the proposed model.

This case shows how to utilize end-cloud computing to enable video transmission with limited spectrum resources. Downsampling the video at the end significantly reduces the data volume and the DL-enabled reconstruction takes full advantage of the abundant computing power of the cloud server.

\begin{figure*}[t]
	\centering
	\begin{subfigure}{0.42\linewidth}
		\centering
		\includegraphics[width=1\linewidth]{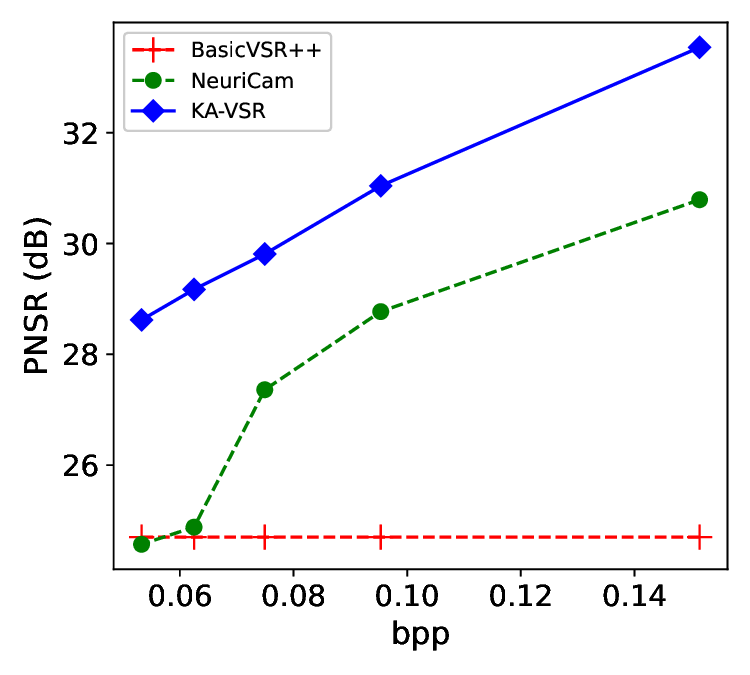}
		\caption{PSNR versus bpp}
		\label{a1}
	\end{subfigure}
	\centering
	\begin{subfigure}{0.42\linewidth}
		\centering
		\includegraphics[width=1\linewidth]{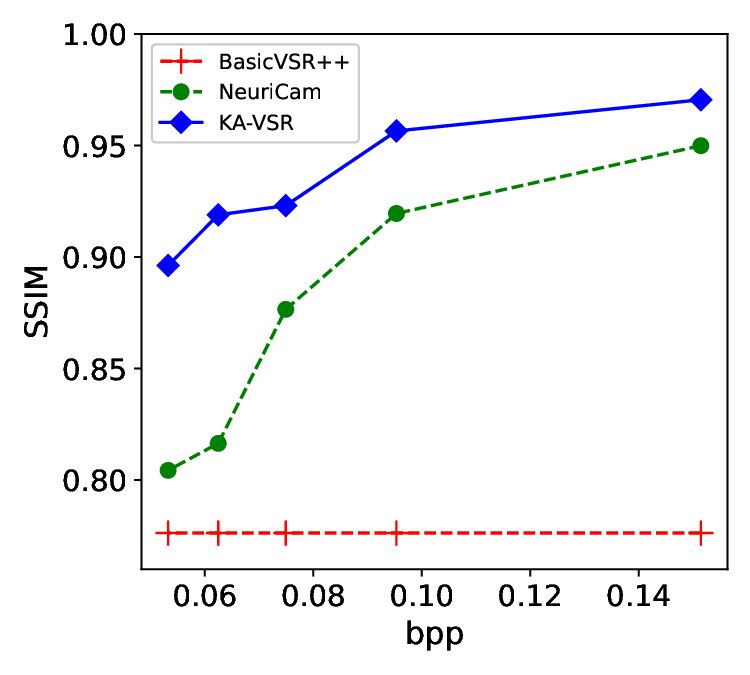}
		\caption{SSIM versus bpp}
		\label{a2}
	\end{subfigure}
	\centering
	\caption{Reconstructed quality under different bit rates, where REDS dataset is used for evaluation. The key frame interval is set to 50, 41, 33, 25, and 15, corresponding to each bpp of videos.}
	\label{case_a}
\end{figure*}

\subsection{Computing Networks enabled Semantic-aware Resource Allocation}
In this study, we design a semantic-aware task offloading system and illustrate that the energy consumption of the local user equipment can be reduced by the proposed system with the optimization of the communication and computation resources. In the proposed semantic-aware task offloading system, several end devices with machine translation tasks are considered, and the semantic features of the sentences are extracted by the semantic encoders at end devices. The end user could execute the task locally using the decoder or offload the tasks to edge servers by transmitting the compressed semantic features. Unlike the conventional edge computing techniques where end devices offload the source data to the edge server, only the extracted semantic information is transmitted to the edge servers in the proposed semantic-aware task offloading system. Therefore, the source data can be compressed and the transmission overhead can be reduced.

However, adopting the semantic scheme represents the improved computation cost, which may affect the battery life of end users. Considering that there is sufficient power supply at the edge server side, the objective is to prolong the battery lifetime of end users, i.e., to minimize the energy consumption of end users. The semantic decoders are deployed at the edge servers to decode the received extracted semantic information. To optimize the resource of the semantic-aware task offloading system and achieve the objective function, the energy consumption for each task should be minimized and the following resources need to be considered. First, end users should decide whether to offload the task to the edge servers or not based on the offloading policy. Meanwhile, end users should determine the computing resource, e.g., the computing frequency, which causes different energy consumption and task execution latency. Additionally, for the task offloading process, the transmit power also needs to be optimized to strike a tradeoff between the energy consumption and the offload latency.

\begin{figure}[t]
\centering
\includegraphics[width=3.25in]{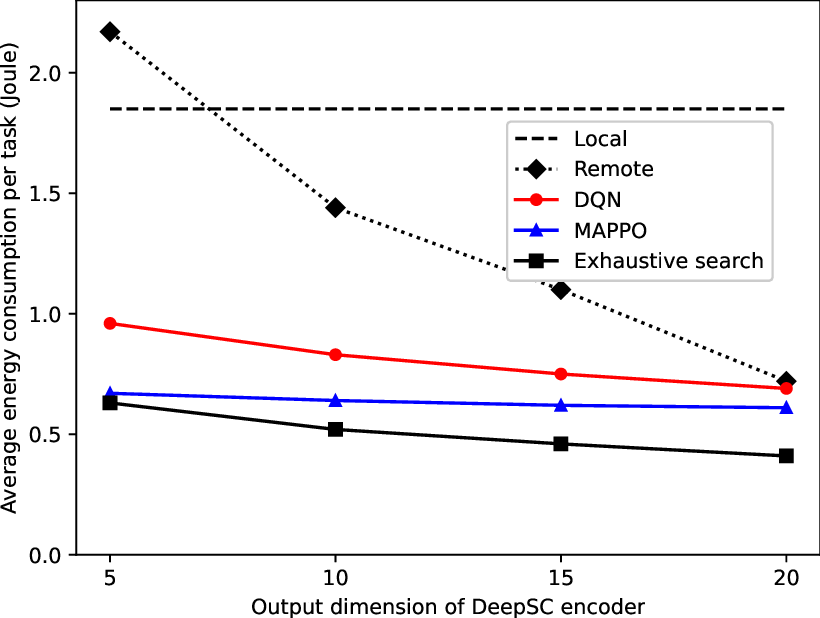}
\caption{Energy consumption over the output dimension of the semantic encoder (number of transmitted symbols per word), where the DeepSC~\cite{DeepSC} model is adopted. The number of end devices is $4$, the transmit power range from $15~\rm{dBm}$ to $24~\rm{dBm}$, and the GPU frequency range from $0.96~\rm{GHz}$ to $1.72~\rm{GHz}$. }
\label{MAPPO_Resource}
\end{figure}

In this study, we propose a multi-agent proximal policy optimization (MAPPO) algorithm~\cite{DeepSC_Resource} to jointly optimize the offloading policy, the computing frequency, and the transmit power of end users to minimize energy consumption. Fig.~\ref{MAPPO_Resource} demonstrates the energy consumption of the proposed semantic-aware task offloading system over the output dimension of the Transformer based DeepSC encoder. It can be observed that the task execution energy consumption can be reduced by leveraging effective resource optimization algorithms. Meanwhile, the MAPPO algorithm is designed to optimize the computing resources and transmission resources jointly and distributedly, achieving lower energy consumption over different output dimensions of the semantic encoders.

\section{Conclusions and Outlook}
In this article, we propose a framework for computing networks enabled semantic communications, which can exploit computing resources in networks to support deep learning enabled semantic communications with demanding computing requirements. Two use cases are provided to validate the effectiveness of the proposed design. To pave the way to computing networks enabled semantic communications, substantial  research is required in the following areas:

\begin{enumerate}
   \item \textbf{Computing capability evaluation:} The existing computing capability metrics are relatively simplistic, such as Floating-point Operations Per Second (FLOPS) and Million Instructions Per Second (MIPS). They are not enough to evaluate the computing capability in real communication systems where communication costs, battery power, and so on should be considered. It is a great challenge to design a good evaluation method for communication systems with computing tasks. With the semantics available in semantic communication, we could design a comprehensive computing resource evaluation framework tailored for different computing tasks.
    \item \textbf{Computing resource sensing:} If we want to update the statuses of the whole computing network, plenty of computing-related information will be retrieved, calculated and transmitted, which requires lots of resources and always loses timely status information. Merging the process of computing resource sensing into communications could be a more efficient way to offer precise status information. One possible approach is to find the mappings between reconstructed/intermediate semantic features and statuses of computing networks.  
    \item \textbf{Theoretical analysis of computing networks enabled semantic communications:} Computing networks and semantic communications are both in their infancy. It is vital to develop analytical theories with mathematical formulation to show the performance limits and direct us to optimize computing networks enabled semantic communication systems. For instance, when the total computing power is fixed, how to allocate the computing power among semantic sampling, semantic coding, and semantic reconstruction models to achieve the best end-to-end performance should be optimized. Measuring semantic information based on the logical probability or age of information might be a promising way to provide a corresponding performance bound.
\end{enumerate}

\bibliographystyle{IEEEtran}
\bibliography{reference}

\end{document}